\documentclass[conference]{IEEEtran}

\ifCLASSINFOpdf
   \usepackage[pdftex]{graphicx}
\else
\fi
\usepackage{pbox}
\usepackage{stfloats}
\usepackage{epstopdf}
\usepackage{amssymb}
\usepackage{graphicx}
\usepackage{pgfplots}
\usepackage{filecontents}
\usepackage{pgfplotstable}
\usepackage{url}
\usetikzlibrary{spy}
\usepackage{amsmath}
\usepackage{cite}
\usepackage{hyperref}
\usepackage{tabularx}
\usepackage{algorithm}
\usepackage{algpseudocode}
\usepackage{threeparttable}
\usepackage{subcaption}
\usepackage{booktabs, caption, makecell}
\usepackage{multirow}
\begin{document}

\title{A Survey of HTTPS Traffic and Services Identification Approaches}

\author{\IEEEauthorblockN{
Wazen M. Shbair \IEEEauthorrefmark{1},
Thibault Cholez \IEEEauthorrefmark{2},
J\'er\^ome Fran\c cois \IEEEauthorrefmark{3},
Isabelle Chrisment \IEEEauthorrefmark{2}
}

\IEEEauthorblockA{
\IEEEauthorrefmark{1} University of Luxembourg, SnT, 29, Avenue J.F Kennedy, L-1855 Luxembourg\\
Email: wazen.shbair@uni.lu\\
\IEEEauthorrefmark{2} University of Lorraine, LORIA, UMR 7503, Vandoeuvre-les-Nancy, F-54506, France\\ 
\IEEEauthorrefmark{3} INRIA Nancy Grand Est, 615 rue du Jardin Botanique, 54600 Villers-les-Nancy, France\\
Email: \{thibault.cholez, jerome.francois, isabelle.chrisment\}@loria.fr\\
 }
  }
  
\maketitle

\begin{abstract}
HTTPS is quickly rising alongside the need of Internet users to benefit from security and privacy when accessing the Web, and it becomes the predominant application protocol on the Internet. This migration towards a secure Web using HTTPS comes with important challenges related to the management of HTTPS traffic to guarantee basic network properties such as security, QoS, reliability, etc. But encryption undermines the effectiveness of standard monitoring techniques and makes it difficult for ISPs and network administrators to properly identify and manage the services behind HTTPS traffic. This survey details the techniques used to monitor HTTPS traffic, from the most basic level of protocol identification (TLS, HTTPS), to the finest identification of precise services. We show that protocol identification is well mastered while more precise levels keep being challenging despite recent advances. We also describe practical solutions that lead us to discuss the trade-off between security and privacy and the research directions to guarantee both of them.


\end{abstract}


\section{Introduction}
The global trend toward an encrypted Web quickly made HTTPS the dominant share of Web traffic \cite{Naylor2014}. According to Cisco 2016 annual security report, statistics show that HTTPS accounts for 57\% of all Web traffic in October 2015 \cite{cisco2016}. That number is in line with ISPs: in Europe, French ISPs reported that the amount of encrypted traffic reached 50\% of the Internet traffic in 2015 \cite{Avis2015} against only 5\% back in 2012, while another ISP based in North America expects 65-70\% of HTTPS traffic by the end of 2016 \cite{sand2016}. There are multiple reasons behind this migration towards HTTPS:
\begin{itemize}
\itemsep0em 
\item The personalization of the Web and the concern of users' privacy and security \cite{Hilts2015}.
\item The development of cloud-based services (online storage, backup-servers, etc.) that hold sensitive data \cite{cisco2016}.
\item The wide spread of mobile applications, which generate inherently encrypted traffic \cite{cisco2016}.
\item The arising of programs such as the Electronic Frontier Foundation's "Let's Encrypt" that facilitate the move toward HTTPS by a free, automated, and open SSL Certificate Authority \cite{sand2016}.
\item The diffusion of high-speed Internet and the development of hardware equipment natively handling encryption to reduce computation overhead \cite{kim2015method}.
\item The arising of Over-The-Top (OTT) content providers that want to keep a maximum level of control over their traffic by obfuscating to other actors \cite{Erman2011}.
\end{itemize} 

On the one hand, moving towards secure web using HTTPS allows users and content providers benefit from better security and privacy. On the other hand, for ISPs and network administrators, encryption makes them "blind" to their network traffic and curtails their capacity to perform proper network management activities, such as traffic engineering, capacity planning, performance/failure monitoring, or caching \cite{haffner2005acas}. For example, HTTPS prevents operators from applying QoS measurements that give a priority to critical services, or to use caching techniques to reduce network latency and congestion. While from the security side, HTTPS makes security monitoring methods unable to understand the traffic and to identify anomalies or malicious activities that can be hidden in encrypted connections \cite{husak2016https}. Bortolameotti et al. \cite{bortolameotti2015indicators} have proposed indicators for malicious HTTPS connections that can be used in \textit{Data Exfiltration} scenario, where a compromised enterprise's machine transfers sensitive information to an external server controlled by an attacker over a HTTPS channel to circumvent the security monitoring techniques. Therefore, there is a high demand for solutions able to analyse HTTPS traffic. 

Previous surveys \cite{Callado2009, Valenti2013, Finsterbusch2014, Velan2015} are a valuable indicator to understand the evolution of the Internet and how the academic and industrial communities handle its traffic classification. Table \ref{published-surveys} summarizes the main traffic classification goal of these published surveys. The most recent one by Velan et al. \cite{Velan2015} is focused on classification approaches for the identification of encryption protocols over the Internet. They show that in the past years, the classification of encrypted traffic in large protocol categories such as IPsec, SSL/TLS, SSH, BitTorrent, Skype, etc. has been widely investigated in the community. They conclude that simply identifying encrypted traffic is not enough but should be improved to identify the underlying protocol. They also state that much work was conducted on SSH while TLS should now be at the center of such studies regarding its importance. In this survey, we take these conclusions and even go deeper by focusing on a single type of underlying protocol, HTTPS, while trying to go further in the identification process to identify precise services. We think that this focus is necessary because of the increased amount and complexity of web applications and services run within HTTPS traffic \cite{Naylor2014, husak2015network, Shbair2016}.

\begin{table*}[hpt]
\caption{The published surveys in the field of traffic classification}
\label{published-surveys}
\begin{center}
\begin{tabular}{|l|c|c|c|}
\hline
\textbf{Survey} &\textbf{Covered Period} & \textbf{Focus}&\textbf{Publication year}\\
\hline
\hline
Callado et al. \cite{Callado2009}&2002-2008&Application identification&2009 \\
\hline
Zhang et al. \cite{Zhang2009} & 1994-2008 & Identify P2P applications&2009 \\
\hline
Valenti et al. \cite{Valenti2013}&2004-2013&Application identification&2013 \\
\hline
Finsterbusch et al. \cite{Finsterbusch2014}&2009-2013 & Payload-based identification &2014 \\
\hline
Velan et al. \cite{Velan2015}&2005-2014 & Encrypted traffic identification&2015\\
\hline
\end{tabular}
\end{center}
\end{table*}

\begin{figure}
\setlength{\belowcaptionskip}{-10pt}
\centering
\includegraphics[scale=0.3]{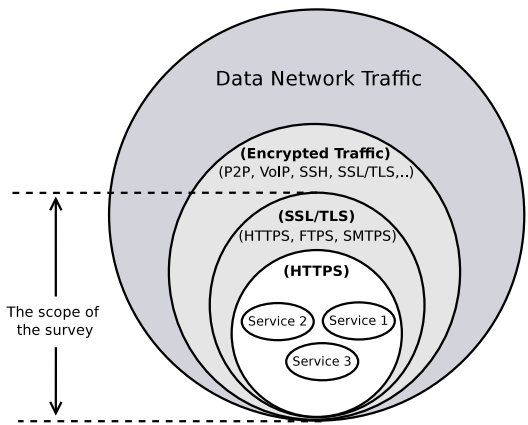}
\caption{Granularity of traffic classification}
\label{scope}
\end{figure}

However, before performing such a deep traffic analysis, the encryption protocol and the top-level application need to be identified first \cite{Cao2013}. In our survey, as illustrated in Figure \ref{scope}, we propose a taxonomy of the related work based on the necessary steps required to identify HTTPS services. We first consider the identification of TLS among other traffic types. Secondly we investigate the methods used to recognize  HTTPS, and we identify the web applications or services in a third time. Our goal is to provide a complete view of HTTPS traffic identification methods that can be used to apply network management, and to identify the remaining research topics in this area.

The remaining part of this paper is organized as follows. The evolution of SSL and TLS protocols and an overview of the protocol structure are presented in Section II. The identification of TLS traffic is explored in Section III, while in Section IV we present the relevant work for recognizing HTTPS traffic. Section V provides a deeper view inside HTTPS traffic, where specific web applications and services are identified. Practical solutions for monitoring and filtering HTTPS traffic are tackled in Section VI. Finally, Section VII concludes the survey with a discussion about the central question of privacy when monitoring encrypted traffic and gives possible research directions.

\section{Background on TLS and HTTPS Protocols}
\subsection{Protocol History}
Nowadays, SSL and TLS protocols are known as the key encrypted protocols over Internet. Those two names come from the protocol history. At the end of 1994 Netscape included the support of the second version of SSL (SSL 2.0) in Netscape Navigator after solving many issues with the first version which was just used inside Netscape Corporation. 
In response, Microsoft also introduced in 1995 an encryption protocol named Private Communication Technology (PCT) that was very close to SSL 2.0 \cite{oppliger2009ssl}. The publication of two concurrent encryption protocols created a lot of confusion in the security community, since applications needed to support both for interoperability reasons. To resolve this issue, the IETF formed a working group in 1996 to standardize a unified TLS protocol. After a long discussion with the related parties, the first version of standard protocol (TLS 1.0) appeared in January 1999. In April 2006, the TLS protocol version 1.1 (TLS 1.1) was released, followed by TLS 1.2 in August 2008 which is specified in the RFC5246 \cite{dierks71rfc} standard. The most recent version of TLS is TLS 1.3, but is still a draft. The new version has the same specifications but with improvements of the encryption algorithms parameters and the handshake \cite{tls132015}. We will use the term “TLS” in the remainder of this survey.


\subsection{Overview of TLS}
\begin{figure}[hbtp]
\setlength{\belowcaptionskip}{-5pt}
\centering
\includegraphics[scale=0.5]{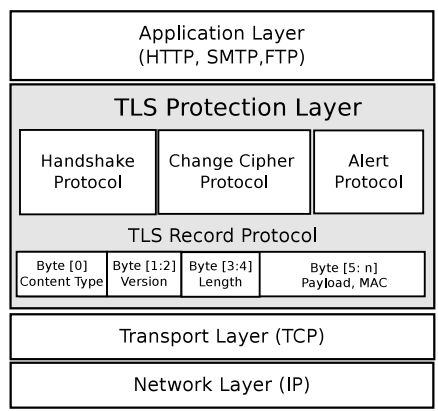}
\caption{The TLS Layers and sub-protocols}
\label{tlssublayer}
\end{figure}

TLS operates below the Application layer and above Transport layer, as illustrated in Figure \ref{tlssublayer}. 
TLS contains two layers, the top-layer holds three protocols: (1) the Handshake protocol is responsible for negotiation to establish or resume a secure connection; (2) the Change Cipher Specification manages the modifications in the ciphering parameters, such as the ciphering algorithm; (3) the SSL Alert Protocol signals problems with SSL connections and allows the communicating peers to exchange alert messages. The lower-layer holds the Record protocol, which can be presented as an envelope for application data and TLS messages from the protocols above. The Record protocol is responsible for splitting data into fragments, which are optionally compressed, authenticated with a Message Authentication Code (MAC), encrypted and finally transmitted \cite{oppliger2009ssl}. 

\begin{figure}[hbtp]
\setlength{\belowcaptionskip}{-5pt}
\centering
\includegraphics[scale=0.35]{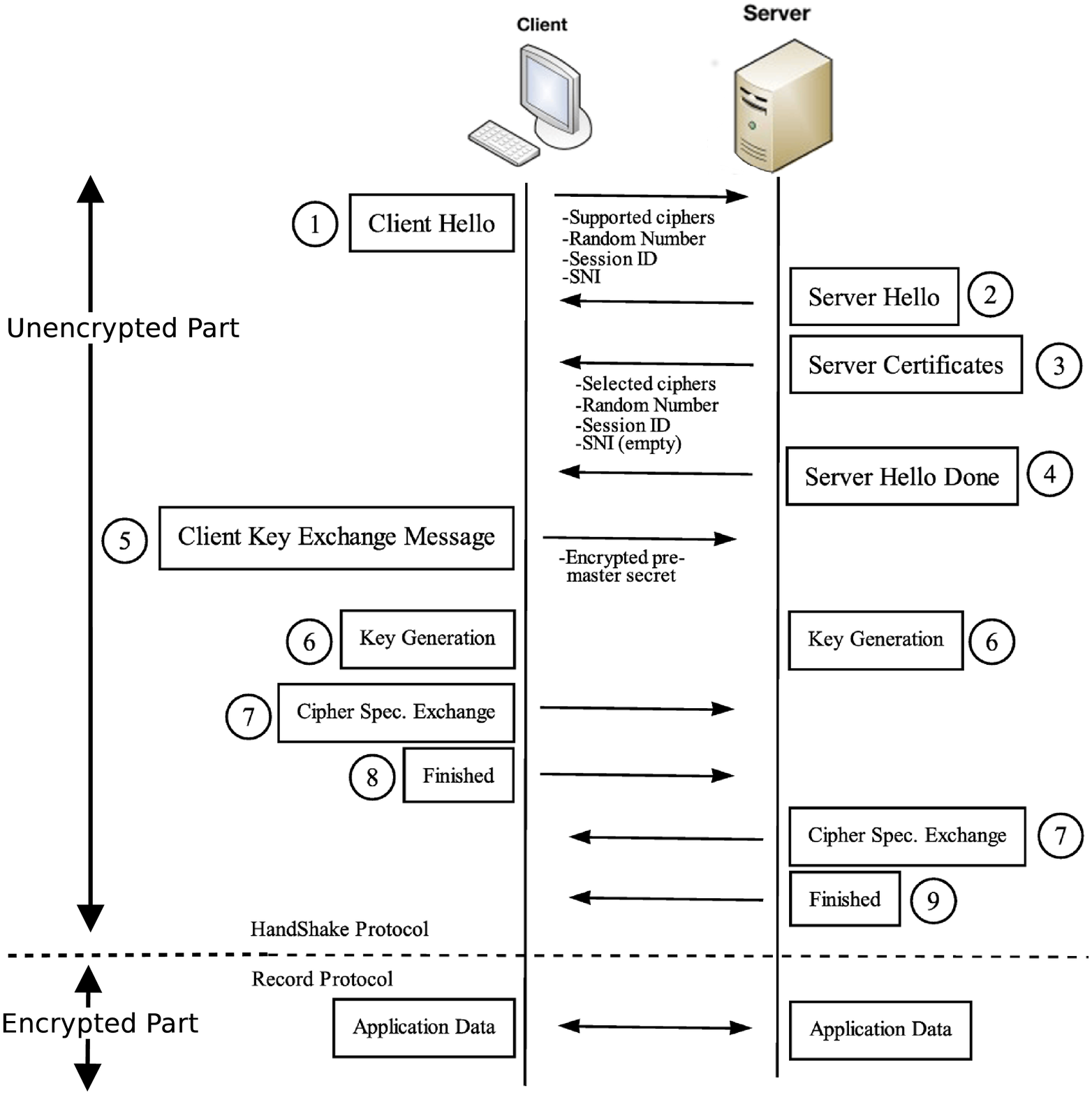}
\caption{The TLS handshake protocol}
\label{tlshandshake}
\end{figure}

The Handshake protocol is of prime importance because it defines the first interactions and is responsible for many configuration aspects such as managing cipher suite negotiation, server/client authentication, and session key exchange. During cipher suite negotiation, client and server 
agree on a cipher suite that will be used to exchange data. 
In authentication, both parties prove their identity using Public Key Infrastructure (PKI) method. In the session key exchange, client and server exchange random and special numbers, called the Pre-Master Secret. These numbers are used to create their Master Secret, which will be used as key for encryption in the phase of exchanging data \cite{tls2016, oppliger2009ssl}. Figure \ref{tlshandshake} details the sequence of TLS Handshake protocol messages:
\begin{enumerate}
\item Client-Hello message contains the usable cipher suites, supported extensions and a random number.
\item Server-Hello message holds the selected cipher, supported extensions and a random number.
\item Server-Certificate message contains a certificate with the server public key.
\item Server-Hello-Done indicates the end of the Server-Hello and associated messages. If the client receives a request for its certificate, it sends a Client-Certificate message. 
\item Based on the server random number, the client generates a random Pre-Master Secret, encrypts it with the public key given in the server's certificate and sends it to the server.
\item Both client and server generate a master secret from the Pre-master secret and exchanged random values. 
\item The client and server exchange "Change cipher spec." to start using the new keys for encryption.
\item The client sends "Client finished" message to verify that the key exchange and authentication processes were successful.
\item Server sends "Server finished" message to the client.
\end{enumerate}
Once both sides have received and validated the Finished message from its peer, they can send and receive application data over the new TLS connection.

\subsection{Introduction to HTTPS protocol}
HTTPS is a protocol providing secure web communications. It simply consists in the Hypertext Transfer Protocol (HTTP) running within a secured TLS connection. To differentiate both services, HTTP uses the port 80 by default while HTTPS uses the port 443. Encrypted-HTTP traffic is still possible without HTTPS but is fundamentally different. A user who needs privacy can still access the desired website over VPN, SSH tunnel, ToR or SSL Proxy. In such case, the website's URL starts with "http://", which means that the web server does not provide a secure connection by itself and that the secure link parameters are configured within the chosen method, but not in the remote server. In HTTPS, the website's URL starts with "https://" and the website is hosted on an authenticated server that owns a SSL certificate. Thus, client and server negotiate the secure connection parameters before exchanging data thanks to HTTP over a dedicated secure link between them \cite{miller2014know}.

\section{Identification of TLS Traffic}
In this section, we review studies that detect TLS protocol's traffic among other types of encrypted traffic. We survey the works providing a Boolean TLS classification method (i.e., TLS vs. non-TLS). This is motivated by the need to recognize the high-level protocol (TLS) before dealing with sub-protocols running within. Later, we will consider that we already have TLS traffic for the next fine-grained level of identification. The relevant work to identify TLS traffic can be grouped into three methods: port-based, protocol-structure based and machine learning based.

\subsection{Port-based method}
Port-based method is a straightforward approach to identify Internet applications and protocols, since the transport layer port numbers are assigned by the Internet Assigned Numbers Authority (IANA). However, to identify TLS traffic effectively it impossible to use only port-number, since the TLS protocol is widely-used with many application layer protocols. For instance, HTTPS, FTPS and SMTPS protocols use TLS over port 443, 990, 465 respectively.
Authors in \cite{Bernaille2007} observe that 8\% of non-TLS traffic use standard TLS ports, while 6.8\% of TLS traffic use ports not associated with TLS. This can be explained by misconfigured web servers or users trying to conceal their activities to avoid port-based filtering \cite{McCarthy2011}. 

\subsection{Protocol-Structure based method}
 Deep Packet Inspection (DPI) technique has been used to identify TLS traffic by examining packets payload with standard format of TLS protocol. The reviewed studies \cite{Bernaille2007, sun2010, liu2012drpsd, kim2015method} share the idea of inspecting packets payload for detecting TLS traffic based on the TLS Record Protocol structure, as shown in Figure \ref{tlssublayer}. The first five Bytes of the TLS Record are: 

\begin{itemize}
\itemsep0em
\item Byte [0] holds content type, as shown in Table \ref{content}.
\item Bytes [1:2] holds protocol version. The values in common use are showed in Table \ref{version}.
\item Bytes [3:4] holds the length of encrypted payload.
\end{itemize}

\begin{table*}
\centering
\caption{TLS identification methods}
\label{tlsidentify}
\begin{tabular}{|l|l|l|c|c|}
\hline
\textbf{Paper} & \textbf{Features} & \multicolumn{1}{c|}{\textbf{Method}} & \textbf{Accuracy} & \textbf{Publish Year} \\ \hline \hline
Bernaille et al. \cite{Bernaille2007} & Port number & \multicolumn{1}{c|}{Port-based} & - & 2007 \\ \hline
Bernaille et al.\cite{Bernaille2007} & Server-Hello packets & \multirow{4}{*}{Protocol-Structure} & 100\% & 2007 \\ \cline{1-2} \cline{4-5} 
Liu et al. \cite{liu2012drpsd} & First 8 Packets &  & 99.17\% & 2012 \\ \cline{1-2} \cline{4-5} 
Finsterbusch et al. \cite{Finsterbusch2014} & DPI &  & 100\% & 2014 \\ \cline{1-2} \cline{4-5} 
Kim et al. \cite{kim2015method} & First 5 bytes &  & 100\% & 2015 \\ \hline
McCarthy et al. \cite{McCarthy2011} & Packet size, timing & Machine learning & 95\% & 2011 \\ \hline
\end{tabular}
\end{table*}

\begin{table}[t]
\setlength{\belowcaptionskip}{-5pt}
    \caption{TLS Record first five bytes contents}
    \begin{subtable}{.5\linewidth}
      \caption{TLS Content Types}
\label{content}
\begin{tabular}{|l|c|}\hline
\textbf{Type}&\textbf{Hex}\\ \hline \hline
ChangeCipherSpec&0xl4 \\ \hline
Alert&0xl5  \\\hline
Handshake&0xl6 \\\hline
Application&0x17 \\ \hline
\end{tabular}
    \end{subtable}%
    \begin{subtable}{.5\linewidth}
    \centering
  	\caption{TLS version}
	\label{version}
	\begin{tabular}{|c|c|}\hline
	\textbf{Version}&\textbf{Hex} \\ \hline \hline
	SSLv3&0x0300 \\ \hline
	TLS 1.0&0x0301 \\ \hline
	TLS 1.1&0x0302 \\\hline
	TLS 1.2&0x0303 \\\hline
	\end{tabular}
 \end{subtable} 
\end{table}

Bernaille et al. \cite{Bernaille2007} are motivated to identify TLS traffic as early as possible, so they use this format to detect Server-Hello packets. As illustrated in Figure \ref{tlshandshake}, the Server-Hello packet is a part of the TLS handshake protocol and it sets the parameters of the TLS connection (e.g., version and encryption algorithm). Therefore, the presence of a valid Server-Hello is a strong indication that the flow is a TLS one. 

Authors in \cite{kim2015method} propose a "TLS Traffic Detector" to isolate pure TLS flows, which are then used in more deep identification to recognize services behind them. The TLS detector compares the first 5 bytes of packets payload (i.e., Bytes [0:4] as explained above) with the TLS record format to take a decision. It benefits from the idea that TLS packet payloads start with the same structure. So checking the first few bytes of the payload for any packets in the flow (not just on the Server-Hello packet as in \cite{Bernaille2007}) is sufficient to mark a flow as TLS. 

Finsterbusch et al. \cite{Finsterbusch2014} evaluate the OpenDPI \footnote{https://github.com/thomasbhatia/OpenDPI} approach that has been used for traffic identification based on DPI. OpenDPI is able to classify TLS traffic with an accuracy of 100\% by using the information in the TLS Record protocol to identify TLS flows in two phases \cite{liu2012drpsd}. In the first phase, it detects a packet which has one Record Protocol Structure in the payload and the payload length is sufficient to read the content type and the TLS version. In the second phase, OpenDPI intercepts the next following packet in the reverse direction, if it has one or more TLS Record protocol structures, then OpenDPI marks that packet as TLS and it continues to check all packets in both directions.

Liu et al. \cite{liu2012drpsd} present a method to detect TLS traffic, named Double Record Protocol Structure Detection (DRPSD). They need to avoid checking all packets to identify TLS traffic. The DRPSD uses only 8 packets to recognize TLS traffic based on detecting the format of TLS Record Protocol. Using a private dataset, the accuracy comparison shows that the OpenDPI achieves 87.69\% accuracy and the DRPSD method has an accuracy of 99.17\%.

\subsection{Machine learning based method}
Machine learning is a type of artificial intelligence that gives computers the ability to learn without being explicitly programmed. The basic requirements are a training dataset (i.e., solved examples), statistical features, algorithms and evaluation techniques. The learning process is divided into three phases; Training, Classification and Validation. In training, the statistical features and machine learning algorithms are trained to make prediction. The output of training phase is a model used in Classification phase to identify unseen data. In Validation phase, the results of classification are validated to measure the performance of the classification model \cite{Valenti2013}. 

Machine learning approach has an important advantage related to its applicability to encrypted traffic. Encryption motivates the usage of this approach to address the limitation of legacy methods (i.e., IP address, port-based, DPI) to identify TLS traffic. Thus, the flow statistics such as flow duration, mean packet size, and mean inter-arrival time are used as features to build a statistical profile for TLS protocol \cite{sun2010novel}. The feasibility of machine learning based method in the context of recognizing TLS traffic was performed in \cite{McCarthy2011}. Four machine learning algorithms (AdaBoost, C4.5, RIPPER and Naive Bayesian) have been evaluated with 22 statistical features (e.g., Mean, Standard deviation, Max, Min, etc.) for the packet size and Inter-Arrival time. The classification results are either Native-TLS, TLS-Tunneled, or Non-TLS. Table \ref{TLS-ML} presents accuracy and False Positive Rate(FPR) of the machine learning algorithm for identifying TLS flows. The AdaBoost algorithm achieves the highest accuracy with 95\% of flows classified correctly as TLS and a 4\% False Positive Rate.

\begin{table}
\setlength{\belowcaptionskip}{-5pt}
\centering
\caption{Machine learning algorithms performance to identify TLS flows \cite{McCarthy2011}}
\label{TLS-ML}
\begin{tabular}{|c|c|c|c|c|}
\hline
& \textbf{AdaBoost}&\textbf{C4.5}&\textbf{Naive Bayes}& \textbf{RIPPER} \\
\hline
\hline
Accuracy & 95.69\% & 85.13\% & 89.26\% & 82.59\% \\
\hline
FPR TLS & 0.04\% & 0.14\% & 0.11\% & 0.17\% \\
\hline
FPR Non-TLS & 0.02\% & 0.01\% & 0.01\% & 0.01\%\\
\hline
\end{tabular}
\end{table}

\subsection{Summary}
To summarize this section, we notice that the identification of TLS is mainly handled by (1) using the TLS record format; 
(2) employing machine learning approach over the encrypted payload, as shown in Table \ref{tlsidentify}. Based on experimental results given in the related work, we are able to recognize TLS traffic among other types of encrypted traffic with a high level of accuracy. Hence, it is possible to detect and identify TLS traffic with high level of confidence and this is no more a research topic. However, investigating protocols run inside TLS is a totally different challenge. In the following section, we overview the usage of TLS protocol for the HTTP application with the goal of predicting exactly which TLS flows hold HTTP traffic.

\section{Identification of HTTPS Traffic}
Among applications that use TLS protocol, HTTP is the most used one \cite{Velan2015}. Hence, in this section we consider the studies addressing the challenge of detecting HTTP traffic inside TLS (i.e., HTTPS). 

\begin{table*}[]
\setlength{\belowcaptionskip}{-5pt}
\centering
\caption{HTTPS identification methods}
\label{HTTPSidentif}
\begin{threeparttable}
\begin{tabular}{|l|l|c|c|c|}
\hline
\textbf{Paper} & \textbf{Features} & \textbf{Method} & \textbf{Accuracy} & \textbf{Publish Year} \\ \hline \hline
\cite{wright2006inferring, Holz2011, Durumeric2013, Hilts2015,bortolameotti2015indicators, Velan2016, Husak2016, Shbair2016} & Port 443& Port based& 100\%*& 2006-2016 \\ \hline
Wright et al. \cite{wright2006inferring} & Packet size, timing, direction & \begin{tabular}[c]{@{}c@{}}Behaviour based\\ (KNN, HMM)\end{tabular} & 100\%, 88\% & 2006 \\ \hline
Haffner et al. \cite{haffner2005acas} & Keywords & \multirow{3}{*}{Machine learning} & 99.2\% & 2005 \\ \cline{1-2} \cline{4-5} 
Bernaille et al.\cite{Bernaille2007} & TLS-Format, First 5 packets size &  & 85\% & 2007 \\ \cline{1-2}
Sun et al. \cite{sun2010novel} & \begin{tabular}[c]{@{}l@{}}TLS-Format, Packets (size, timing)\\ flow duration, Packets number\end{tabular} &  & 93.13\% & 2010 \\ \hline
\end{tabular}
\begin{tablenotes}\footnotesize
\item[*]  If user not malicious (i.e., alter port number)
\end{tablenotes}
\end{threeparttable}
\end{table*} 

\subsection{Port-based method}
Basically, we can use the port-number 443 to identify HTTPS traffic, but port 443 can also be used by malicious applications to hide their activities behind the HTTPS port to give an indication that a Web browsing traffic is running \cite{McCarthy2011}. Alternatively, some HTTPS Web server can be configured to use a different port number \cite{Bernaille2007}. 
Many approaches were proposed to overcome the usage of non-standard port with HTTPS, ranging from behaviour based method to machine learning ones as described below. 
In spite of that, port 443 is widely used in the large body of literature \cite{wright2006inferring, Holz2011, Durumeric2013, Hilts2015,bortolameotti2015indicators, Velan2016, Husak2016, Shbair2016} to collect and build HTTPS dataset for further experiments. 
\subsection{Behaviour based method}
Wright et al. \cite{wright2006inferring} demonstrate how application behaviour still can be used as a signature to identify the application, even if its traffic is transmitted via HTTPS flows. They use the fact that some information remains intact after encryption like packet size, timing, and direction to identify the common application protocols by using k-Nearest Neighbor (kNN) and Hidden Markov Model (HMM). The KNN detects HTTPS flows with 100\% accuracy and the HMM performs 88\% accuracy.

\subsection{Machine learning based}
Haffner et al. \cite{haffner2005acas} propose extracting statistical signature from the packet payload. In the case of unencrypted traffic they extract ASCII words from the data stream as features. But for HTTPS they extract words from the handshake phase, since it is unencrypted as shown in Figure \ref{tlshandshake}. The existence and the location of such words in the first 64-Bytes of a reassembled TCP data stream is encoded in binary vector and used as input for different machine learning algorithms (Naive Bayes, AdaBoost and Maximum Entropy). The evaluation over dataset from ISP shows that AdaBoost identifies HTTPS traffic with 99.2\% accuracy.

Authors in \cite{Bernaille2007, sun2010novel} share the concept of identifying HTTPS in two steps; in the first step TLS traffic is detected based on the protocol-format as discussed in section III-B, while in the second step HTTP traffic in TLS channel is recognized using machine learning method. In \cite{bernaille2006traffic}, authors use the size of first five packets of a TCP connection to identify HTTPS application with 81.8\% accuracy rate. The performance of their classifier has been improved (up to 85\%) in \cite{Bernaille2007} by adding a pre-step phase, where they first detect TLS traffic based on protocol-format and then identify the HTTP traffic within TLS. Sun et al. \cite{sun2010novel} propose a hybrid solution, which firstly detects TLS protocol by inspecting TLS protocol-format, then apply a machine learning algorithm to determine application protocols run with TLS connection. The Naive Bayes algorithm is used with 8 statistical features; Mean, Maximum, Minimum of packet length, and Mean, Maximum, Minimum of Inter-Arrival time, flow duration and number of packets. Using a private dataset, results show the ability to recognize over 99\% of TLS traffic and to detect HTTPS traffic with 93.13\% accuracy.


\subsection{Summary}
The related work in the identification of HTTP application within TLS protocol, as illustrated in Table \ref{HTTPSidentif}, used different methods with an acceptable level accuracy but each one with its own built dataset that prevent others from having a strong and fair comparison of their respective identification accuracy. The situation will remain ambiguous in the absence of a reference dataset for all. That leads to another research question about reproducible-research and dataset construction \cite{Velan2015}. We should also question the representativity of dataset, since HTTPS nowadays is a multi-purpose protocol (i.e., it can deliver video, music, games, etc.). The next section delves more into HTTPS application traffic itself, to identify the work and the perspectives to name the specific web applications and services that generate HTTPS traffic.

\section{Identification of services inside HTTPS}
This section explores the methods, which precisely identify the real source of HTTPS (i.e., Web services). The increased complexity of web applications provides the ability to deliver very different kinds of services such as email, games, online storage, content providers, maps, social media plug-in, etc., all transmitted via HTTPS flows \cite{kim2015method, Shbair2016}. The identification of HTTPS services is a serious challenge, since most of the legacy techniques like DPI lose their power when facing encryption.

\subsection{Website fingerprinting method}
Identifying the accessed websites over secure connections is well-known as \textit{Website Fingerprinting}. This method is presented in most of relevant work \cite{Bissias2005, Herrmann2009, Chang2010, Pironti2012}. Cheng et al. \cite{Cheng1998} propose one of the earliest method to identify the pages visited by users over TLS connection by inspecting TCP/IP header, which contains the size of payload and other information. Their technique is based on calculating the size of a page downloaded to browser, which is often unique among all files in a given site. Moreover, as HTML files cannot be transmitted concurrently with other files, they thus remain distinguishable. At that time (i.e., 1998 and before) the browsers did not use the HTTP Pipelining, which now hides the objects size and order. Miller et al. \cite{miller2014know} proposes a method to identify the accessed page among 500 pages hosted at the same HTTPS website based on clustering techniques to identify patterns in traffic. Their results show the possibility to identify individual pages from the same website accessed over HTTPS with 89\% accuracy. They successfully identify the home-page or internal-pages from a website but at the cost of a specific learning at a single website page level, while more effort is needed to identity embedded services in web pages.

\subsection{Behaviour based method}
In \cite{Schatzmann2010}, the authors develop a passive approach for webmail traffic identification in HTTPS in order to understand the shifting usage trend and mail traffic evolution. Three novel features are proposed (1) service proximity: the presence of POP, IMAP or SMTP server within a domain is a strong indication that a mail sever exists; (2) activity profiles: mail system clients access their e-mail frequently in a scheduled manner, so its possible to build daily and weekly profile to such behaviour; (3) periodicity: the usage of application timers like AJAX technology to periodically (e.g., 5 minutes) check for new messages creates high frequency time pattern and gives strong indication the email service is running. These features are used with Support Vector Machine (SVM) algorithm to differentiate between mail and non-mail services within HTTPS flows. The evaluation over dataset from ISP shows the ability to identify HTTPS mail server with 93.2\% accuracy.

Chen et al. \cite{chen2010side} use the traffic pattern of the AutoComplete function, which populates a list of suggested content with each letter an user enters, such as Google and Yahoo search engine. Despite HTTPS, this small amount of input data causes state transitions in a web application, which can be used to enumerate all possible inputs to match the triggered traffic pattern. Based on real scenarios, they show how such a method can be applied to leak out sensitive information (e.g., Search Keywords) from top online web applications. V. Berg. \cite{maps2011} develops a tool that uses encrypted traffic patterns to identify user activities over Google maps that already runs over HTTPS. The tool collects satellite map tiles and builds a database of the image sizes correlated with their (x,y,z)-triplets coordinates. To identify the accessed region over Google maps, the tool maps the size of images in HTTPS flow to (x,y,z)-triplets and then clusters the results into a specific region. As a proof of concept, the tool's dataset has been configured with city profiles, where it can correctly detect the transition between such cities.

\subsection{SSL certificate based method}
SSL certificates, which are originally used to verify the identities of servers and clients, are also used to recognize the accessed service over HTTPS flows. Kim et al. \cite{kim2015method} use the certificate public information to build SSL/TLS Identification Method (SSIM) to name the services behind HTTPS traffic. The proposed method consists of three modules: (1) TLS Traffic Detector module isolates pure TLS traffic before beginning the service identification; (2) service signature module extracts Certificate authority information, Server IP and Session ID from a SSL certificate; (3) Session ID-IP-based Service Identifier module recognizes non-identified flows from the previous modules by finding relation between server IP and session ID. Based on their experiments, they can classify 95\% of TLS traffic belonging to Google, Facebook and Kakaotalk with about 90\% accuracy for the corresponding services. Authors in \cite{bortolameotti2015indicators} use the information in SSL certificate like certificate validity, release dates and the content of subject alternative name as features to detect suspicious TLS traffic.

\subsection{Protocol-Structure based method}
One widely used technique to identify HTTPS service is based on inspecting a field of TLS protocol header, namely Server Name Indication (SNI), which has been recently implemented in many firewall solutions and content filtering solutions. The SNI is mainly used to allow a client to specify the server hostname when the TLS negotiation starts, as shown in Figure \ref{tlshandshake}. The idea is using SNI to filter HTTPS traffic, since it indicates the name of the remote service a client intends to access. SNI provides the identification system with the power to early abort the access to prohibited services.
 
Bortolameotti et al. \cite{bortolameotti2015indicators} used both SNI and SSL certificate information to detect malicious TLS connections by examining (1) Levenshtein distance between the SNI and top 100 most visited websites; (2) the structure of the server-name string in SNI; (3) the format of the server-name string, which is a DNS hostname format. In \cite{wazen2015}, authors evaluate the reliability of identifying HTTPS based on SNI. They found two inherent weaknesses, regarding (1) backward compatibility and (2) multiple services using a single certificate, which can be used by a client to cheat the identification system. As proof of concept, they develop a web browser plug-in to demonstrate how these weaknesses can be practically used to bypass firewalls relying on SNI to identify HTTPS traffic.

\subsection{Machine learning based method}
Recent work \cite{Shbair2016} argues that the page-level identification is too fine-grained (i.e., Website Fingerprinting), specially in the case of identifying content that is dynamically included in other web pages (video, maps, etc.) Thus, they propose a method for identifying HTTPS at service-level thanks to a multi-level framework, without relying on specific header fields, such as SNI that can be easily altered or the TLS certificate information. The proposed framework uses  machine learning algorithms (Randomforest and C4.5) with a statistical profile library of intended HTTPS services. The profile contains statistical measurements (Mean, variance, Max, Min, etc.) over packets size and the inter-arrival-time over TLS flow packets. For evaluation, real traffic collected from their university network (i.e., private dataset) has been used. Their multi-level framework can identify HTTPS web services with 93\% accuracy.
  
\begin{table*}[]
\setlength{\belowcaptionskip}{-5pt}
\centering
\caption{Services identification inside HTTPS}
\label{httpsservices}
\begin{threeparttable}
\begin{tabular}{|l|l|l|c|c|c|}
\hline
\textbf{Paper} & \textbf{Features} & \textbf{Method} & \textbf{Level of Identification} & \textbf{Accuracy} & \multicolumn{1}{c|}{ \textbf{Publish Year}} \\ \hline \hline
Cheng et al. \cite{Cheng1998} & Packet size and order & \multirow{2}{*}{Website Fingerprinting} & Internal pages & 96\% & \multicolumn{1}{c|}{1998} \\ \cline{1-2} \cline{4-6}
Miller et al.\cite{miller2014know} & Packets size and direction &  & Internal pages & 89\% & 2014 \\ \hline
Schatzmann et al.\cite{Schatzmann2010} & \begin{tabular}[c]{@{}l@{}}Service proximity, activity profiles, \\ session duration, and periodicity\end{tabular} & \multirow{3}{*}{Behaviour based} & Email services & 94.8\% & 2010 \\ \cline{1-2} \cline{4-6} 
Chen et al. \cite{chen2010side} & AutoComplete function traffic &  & Search keywords & - & 2010 \\ \cline{1-2} \cline{4-6} 
Vincent Berg \cite{maps2011} & Images size &  & Google maps activities & - & 2011 \\ \hline
Kim et al. \cite{kim2015method} & Certificate information & \multirow{2}{*}{SSL certificate} & Services & 90\% & 2015 \\ \cline{1-2} \cline{4-6} 
Bortolameotti et al. \cite{bortolameotti2015indicators} & Certificate information &  & Services & 100\%* & 2016 \\ \hline
Shbair et al. \cite{wazen2015} & SNI & \multirow{2}{*}{Protocol Structure} & Services & 100\%* & \multicolumn{1}{c|}{2015} \\ \cline{1-2} \cline{4-6} 
Bortolameotti et al. \cite{bortolameotti2015indicators} & SNI &  & Malicious connections & 100\%* & \multicolumn{1}{c|}{2015} \\ \hline
Shbair et al. \cite{Shbair2016} & SNI, packets size and timing profile & Machine learning & Services & 93\% & \multicolumn{1}{c|}{2016} \\ \hline
\end{tabular}
\begin{tablenotes}\footnotesize
\item[*]  If user not malicious (i.e., alter SNI)
\end{tablenotes}
\end{threeparttable}
\end{table*} 

\subsection{Summary}
Many recent approaches, as summarized in Table \ref{httpsservices}, intend to identify HTTPS services based on the plain-text information that appears in the TLS handshake phase or based on the statistical signature of HTTPS web services. However, the reliability of handshake information for identification still need improvement, as discussed with SNI and SSL certificate. While the reliability of machine learning method has a challenge with the increased complexity of web applications that can be easily extended with new functionalities that may change the application behaviour and the statistical-signature. This complexity creates an overhead to the machine learning based identification methods to re-evaluate their statistical features and re-train classification models regularly to keep their methods effective with updated changes. 

The related work that can precisely name the service behind HTTPS flow are mainly depends on the offline/passive analysis where full HTTPS flows available for training and classification. However, the offline analysis is less critical to the training time duration, computation overhead and classification error over time. That opens a research question about the real-time identification of HTTPS services. The existence of real-time detection will help ISPs and administrators to manage HTTPS services at the right time, and perform the proper network management activities. The side question is to know how currently the HTTPS traffic is monitored and filtered. The answer to this question is presented in the next section.

\section{Practical HTTPS Monitoring and Filtering}
Once HTTPS traffic is identified then it can be monitored and filtered. Monitoring is defined as recording all activities on a network (URLs visited, session duration, bandwidth, etc.) and generating a report that can be used for network management. Some of current HTTPS monitoring approaches use similar techniques than research papers. For example, some monitoring approaches rely on the information exchanged during handshake such as: SSL certificates \cite{Holz2011, Durumeric2013}, the handshake interactions sequence \cite{Korczynski2014} and the most recent is SNI-based monitoring \cite{wazen2015}, which has been implemented in solutions as Clavister Web Content Filtering and Sophos Unified Threat Management (UTM) to monitor accessed websites over HTTPS flows \cite{wazen2015}. HTTPS filtering is intended to restrict access by intercepting the transmitted date between a client and server \cite{McGinnes2010}. Hence, to deploy HTTPS filtering two approaches have been presented: HTTPS proxy server and acquiring TLS encryption keys.


\subsection{HTTPS Proxy Server}
\begin{figure}[hbtp]
\setlength{\belowcaptionskip}{-5pt}
\centering
\includegraphics[scale=0.45]{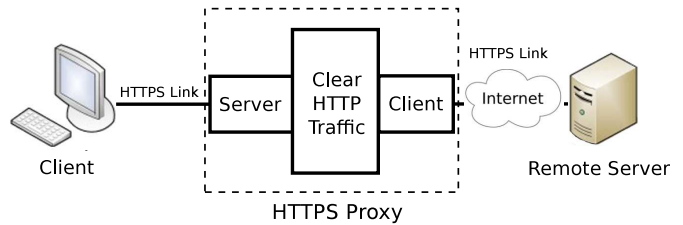}
\caption{HTTPS Proxy Server}
\label{httpsproxy}
\end{figure}

A proxy server is a server acts as a man in the middle to processes and forwards clients requests towards servers. However, when HTTPS is used the proxy server cannot directly access data transmitted via HTTPS, so it pretends to be the intended remote server then it establishes a secure connection to the real server. As shown in Figure \ref{httpsproxy}, when a client connects to the remote server via a HTTPS proxy the client connects to the proxy server, which  plays the role of a destination server by providing its own SSL certificate. Then the proxy establishes another secure connection with the real remote server. By this method all encrypted web traffic is open to the proxy in clear at the expense of users' privacy \cite{wazen2015}.

Existing commercial solutions such as Forefront Threat Management Gateway (TMG) 2010 uses the HTTPS proxy method for HTTPS inspection, which acts as a trusted man-in-the-middle instead of just tunnelling HTTPS connection blindly \cite{TMG2010}. Also the FireEye product uses the proxy model to provide visibility into untrusted TLS traffic. The product is designed to intercept and forward all desired network traffic for temporarily decrypting, examining and then re-encrypting TLS sessions again. The FireEye argues this method responds to the growing number of cyber criminals that use TLS as a cover to get inside organizations and persist undetected \cite{fireeye2015}.

\subsection{Acquiring TLS encryption key}
There are at least two methods for acquiring the decryption keys, the Key-Recovery mechanism and the cracking of  encryption algorithms. In \cite{Abelson1997} author describe the Key-Recovery mechanism or "Key escrow", where all encryption keys are stored in a trusted third party, such as government, or designated private entities. The third party has the right to access keys for authorized law enforcement purpose. As a result, a government may limit access to HTTPS websites that refuse sharing their TLS keys with the escrow system \cite{McGinnes2010}. 
However, cracking encryption algorithms is different from the preceding ones, as it needs high computation power to be able to crack the encryption. A method for cracking is using a flaw in the mathematical algorithm used to encrypt data, such as the factorization of widely used public-key cryptosystems. For instance, RSA 768-bit can be broken with a state of the art algorithm and a high computation power \cite{kleinjung2010factorization}. Adrian et al. \cite{adrian2015imperfect} evaluated the security of Diffie-Hellman key exchange, where they found that 82\% of vulnerable servers use a single 512-bit group, which makes it possible to compromise connections of 7\% of Alexa Top Million HTTPS sites. 

 
\section{Conclusion}
HTTPS is quickly becoming the predominant application protocol on the Internet. It answers to the need of Internet users to benefit from security and privacy when accessing the Web. But the 
increasing amount of HTTPS traffic comes with challenges related to its management to guarantee basic network properties such as security, QoS, reliability, etc. The encryption undermines the effectiveness of standard monitoring techniques and makes it difficult for ISPs and network administrators to properly identify services behind HTTPS traffic and to properly apply network management operations.

This survey provides a focused view of HTTPS traffic identification method, starting from the  identification of the lower-level TLS protocol to the precise identification of HTTPS services. We have found that efficient methods exploiting the standard structures of the TLS protocol are able to identify TLS traffic among other types with a high level of accuracy and are no more a research topic. The identification of HTTPS uses different methods with an acceptable level of detection rate, but the confusing matter is the representativity and the diversity of dataset, which prevent any strong and direct comparison of their respective identification accuracy. This leads to important research questions about reproducible research and the best practices for HTTPS dataset construction and diffusion. Finally, based on the most recent works, we show that some efforts were made to discriminate between services running within HTTPS traffic. While the results are encouraging, most solutions still suffer from significant drawbacks ranging from the specialized identification of a precise application (webmail, maps, etc.) to the inability to operate in real time.

In the previous section of this survey, we noticed that there is a very efficient method to monitor and control HTTPS traffic based on HTTPS proxy. This easy solution has many supporters from National Security Agencies \cite{anssi} to security companies \cite{fireeye2015}, and is even discussed for future Internet technical standards \cite{httpbis_proxy}. However, such a method cannot be treated lightly as it denies the right for privacy for the sake of traffic inspection and therefore creates a paradox between the need for security and users' privacy. The answer to this conflict is not easy, as both sides may have valuable arguments. After the Snowden affair, this issue is known as "Dilemmas of the Internet age"  and has been discussed not only in the academic community but in the overall society and human rights space. Authors in \cite{caloyannides2004privacy} 
claim that large-scale monitoring is ineffective as it is only able to identify trivial crimes, but cannot recognize professional criminals or persons well educated in working under surveillance. Another important question is how 
 to guarantee that an administration in power will never abuse the intercepted information to intimidate its opponents. Authors conclude that if online monitoring may fix some problems, it can create even more serious ones. In another domain, even if enterprise owners may have good arguments for monitoring and filtering access to their network for security, productivity or responsibility reasons \cite{currentware}, it may not be sufficient to legitimate the exposition of employees private data with HTTPS proxy.

In conclusion, while challenging, we think that the identification of HTTPS services without decryption is the way to go to provide a viable compromise between the needed network knowledge to ensure proper management and users' privacy. The research community should focus on proposing new identification techniques offering both security and privacy.




\bibliographystyle{IEEEtran}
\bibliography{ref}
%
\end{document}